\documentstyle[12pt]{article}
\begin{document}

\author{Victor Novozhilov and Yuri Novozhilov  \\
V.A.Fock Department of Theoretical Physics, \\ St.Petersburg State University,
198904, St.Petersburg, Russia }

\title{Collective variables and composite fields
\footnote{to be published in the Proceedings of the Conference "Probability
and Irreversibility in Quantum Mechanics", 5-9 July 1999,
Fondation des Treilles, France\\
This work was supported in part by RFBR (Grant 97-01-01186) and
by GRACENAS (Grant 97-0-14.1-61).
}}
\date{}
\maketitle
\abstract
{
We consider use of collective variables for description of composite fields
as collective phenomena due to the strong coupling regime.
We discuss two approaches, where identification of collective variables
of complex quantum system does not depend on knowledge of other degrees
of freedom: (a) collective variables as parameters of group transformations
changing the path integral of the system, and (b) collective variables
as background fields for quantum system. In the case (a) we briefly present
an approach. In the case (b) we consider fermions in an external scalar
field, which serves as a collective variable in a nonlinear model for
composite scalar field with a finite compositeness scale.
}

{\bf Introduction}
\vskip 3pt

In absence of particle-creation interaction, to describe a composite
field means to solve the Schroedinger equation for constituent particles.
When interaction changes particle number, one should work in the Fock space
, and if coupling is strong , even a single particle should be described in
the Fock space by the column vector with many rows; a single particle
becomes ''dressed''. A composite particle state in the Fock space will
include rows corresponding to different ways of combining indefinite many
more elementary systems into states with the same internal quantum numbers.
In Quantum Field Theory, this situation can be also expressed in language of
the Bethe-Salpeter \cite{Bethe-Salpeter} and the Schwinger-Dyson \cite{Dyson}
equations as necessity to take
into consideration more and more graphs. In other words, in the strong
coupling regime a composite field becomes a collective phenomenon. Its
properties may be quite different from what can be guessed from perturbation
theory.

As a collective phenomenon, composite particles in a strong coupling regime
- with respect to the underlying fundamental particles - are similar to
macroscopic systems with respect to underlying microscopic systems.
Bogolyubov description of superconductivity in terms of collective variables
belonging to the microscopic level is a well-known example of such an
approach \cite{NN}. In complex systems, identification of collective degrees
of freedom is a starting point for their description.

In QFT, to find an appropriate picture of composite particles in a strong
coupling regime is one of the main problems of the theory. In Quantum
Chromodynamics a solution of this problem is needed in order to compare low
energy QCD with an experiment.Understanding of composite quantum fields is
necessary in order to look for properties of theories at different mass
scales.

In this paper we consider two approaches for choosing collective variables
in QFT, namely, (a) collective variables as parameters of non-invariance
groups, and (b) collective variables as external fields' background for
quantum fields.

\vskip 3pt
{\bf Quantum composite fields as parameters of non-invariance groups}
\vskip 3pt

In the modern QFT collective variables can be introduced in the generating
functional. Consider a field $\Phi (x)$ with the Lagrangian $L(x)$ and
vacuum functional

$$
Z=\int {\it D}\Phi e^{i\int Ldx} \eqno(1)
$$

Let $\Pi (x)$ be a local field with quantum numbers of a composite system.
made out of $\Phi $ .The functional measure ${\it D}\Phi $ is over all
independent degrees of freedom of $\Phi $; thus it includes also degrees of
freedom of $\Pi (x)$ together with remaining degrees $X$ which are
considered inessential. If we change variables

$$
\{\Phi \}\rightarrow \{\Pi ,X\}  \eqno(2)
$$

and transform $Z$

$$
Z=\int {\it D}\Pi {\it D}XJ\left( \partial \Phi /\partial \Pi ,\partial \Phi
/\partial X\right) e^{i\int L\left( \Pi ,X\right) dx}\equiv  $$
$$
Z_{inv}\int {\it
D}\Pi e^{i\int L_{eff}\left( \Pi \right) dx} \eqno(3)
$$

then after integrating out inessential variables $X$ we arrive at the
effective Lagrangian $L_{eff}(\Pi )$ for a collective variable $\Pi $
describing a composite field. The functional $Z_{inv}$ does not depend on $
\Pi $ ; $J$ is a Jacobian of transformation $\{\Phi \}\rightarrow \{\Pi ,X\}$
.

In practice, to find directly variables $X$ and the Jacobian is a difficult
task. However, there are some classes of collective variables $\Pi $, when
knowledge of $X$ is not necessary in order to find an effective Lagrangian $
L_{eff}\left( \Pi \right) $ , namely, when $\Pi $ are parameters of
non-invariance groups and different classes correspond to different groups.
''Non-invariance'' is understood in relation to the vacuum functional : $
\delta Z/\delta \Pi \neq 0$ . In this case one can hope to integrate $\delta
Z/\delta \Pi $ up to a factor in vacuum functional $Z$ depending only on $
\Pi $ .

Consider a group $G$ of transformations $U=\exp i\Pi $ of field $\Phi $, $
\Phi \rightarrow U\Phi \equiv \Phi ^U$ with the invariant measure ${\it D}
(UU^{\prime })={\it D}U$ and the vacuum functional
$$
Z\left[ U\right] =\int {\it D}\Phi \exp \left( i\int dxL\left( \Phi
^U\right) \right) \eqno(4)
$$

Integrating over $U$ we get $G $ - invariants

$$
Z_0=\int {\it D}UZ\left[ U\right] ,Z_{inv}^{-1}=\int {\it D}UZ^{-1}\left[
U\right] \eqno(5)
$$
which can be used in order to substract from $Z$  a  $G$-invariant part
leaving a functional $Z_U$ for $U$

$$
ZZ_0^{-1}\simeq ZZ_{inv}^{-1}=Z_U\equiv \int {\it D}U\exp \left( i\int
dxL_{eff}\left( U\right) \right) \eqno(6)
$$

and identifying an effective action for $U=\exp i\Pi $ as

$$
W_{eff}\left( U\right) =\int dxL_{eff}\left( U\right) =-i\ln \frac Z{Z\left[
U\right] } \eqno(7)
$$

We use $\simeq $ to show that $Z_0$ and $Z_{inv}$ differ by $G$ -invariant
terms. Thus, $L_{eff}$ is defined up to $U$ -independent terms.

We see that inessential variables $X$ were effectively integrated out in
integration over all initial degrees of freedom. A replacement $\Phi
\rightarrow \Phi ^U$ in both the measure and lagrangian in $Z$ is just a
change of notations and cannot change $Z.$ Thus , the Jacobian $J$ in (3)
is expressed in terms of $W_{eff}\left( U\right) $.

This is a group-theoretical way to derive low energy effective lagrangians
for chiral field $U=\exp i\Pi $ \cite{AAA+YuVN}, where pseudoscalar field $\Pi $ describes
pions and kaons, an extended chiral field ${\cal U}$ of the extended chiral
group $E\chi $ describing pseudoscalar mesons and diquarks \cite{diquark} , and dilaton
field $\sigma (x)$ related to the conformal transformation $g_{\mu \nu
}\rightarrow (\exp 2\sigma )g_{\mu \nu }$  \cite{PhL}. In this way we get quantum
composite fields $U,{\cal U}$ and $\sigma $ with their vacuum functionals,
i.e. with integration over these fields.

\vskip 6pt
{\center {\bf Quantum fields in classical background and induced classical
action for composite field}
\vskip 3pt
{\bf Induced classical gravity.}}
\vskip 3pt

Quantum matter fields on curved space represent the most general example of
such class. Vacuum fluctuations of matter fields transform a given metric
into a classical dynamical gravitational field with the Einstein action in
the longwave region which got the name of 'Induced Gravity'. This fact was
discovered by Zeldovich and Sakharov \cite{Zeld+Sakharov} and developments
were reviewed by
Adler \cite{Adler}. Due to ambiguity \cite{David} in definition of Newton constant
Induced Gravity was forgotten for a decade and reconsidered \cite{Grav} with modern
calculational tools in the framework of bosonization approach.
''Einsteinization'' \cite{Einst} of external metric under influence of collective
vacuum mode of matter fields represents a phenomenon of 'partial
compositeness', when quantum fields do not create a new field, but only
provide an existing field with kinetic term and interactions.

Let $L\left( \Psi ,g_{\mu \nu }\right) $ be a lagrangian of matter fields $
\Psi $ on curved background $g_{\mu \nu }$ and $Z_\Psi \left[ g_{\mu \nu
}\right] $ denote the vacuum functional

$$
Z_\Psi \left[ g_{\mu \nu }\right] =\int {\it D}\Psi \exp \left( \int
dxL\left( \Psi ,g_{\mu \nu }\right) \right) \eqno(8)
$$

Integration over matter $\Psi $ in $Z_\Psi $ takes into account vacuum
fluctuations of fields $\Psi $. As a result, an external metric $g_{\mu \nu
} $ entering the lagrangian $L$ in integrand comes out as an argument of an
effective action $W\left( g_{\mu \nu }\right) $

$$
\exp \left( iW\left( g_{\mu \nu }\right) \right) =Z_\Psi \left[ g_{\mu \nu
}\right]     \eqno(9)
$$

It is the action $W\left( g_{\mu \nu }\right) $ that describes induced
Einsteinian gravity if evaluation of $Z_\Psi $ is implemented carefully.

\vskip 3pt
{\bf Induced action for mesons with finite compositeness scale}
\vskip 6pt

As an example, consider quantum quark field $\psi $ in background of scalar
field $S\left( x\right) $ which serves as a collective variable in  a model
of scalar mesons \cite{Fock} . Model of flavor vector mesons in a similar approach
was constructed earlier \cite{VYuN}.

The purpose of the model is to develop a theory of isotriplet and isosinglet
scalar mesons $a_0$ and $f_0$ from the viewpoint of the low energy QCD as a
part of renormalized theory. Induced action for scalar mesons is obtained by
integration over quark fields in the quark functional integral. But  we
add an assumption that these mesons exist within finite compositeness region
with the upper scale $\Lambda_s$ , thereby describing the slope of the coupling
constant beyond the chiral breaking scale.
 We use an approach proposed for the model of
composite electroweak bosons \cite{YuVN} and definition of a composite particle in
renormalisation theory through radiative corrections \cite{Hayashi}. We use also
similarity between low energy scalars and Higgs fields when gauge fields are
also present.

Integration over fermions in the quark path integral in its simplest form is
equivalent to one-loop approximation which produces polynomials of scalar
fields and,consequently, may lead to scalar condensate. Advantage of the
path integral approach is that it can easily take into account the quark and
gluon condensates. The problem which arises in the low energy QCD is whether
the compositeness scale $\Lambda _s$ for scalars coincides with the scale of
the chiral symmetry breaking or it should be introduced as a separate low
energy parameter. We assume that $\Lambda _s$ coincides with the upper scale
of the QCD low energy region defined in terms of the quark and gluon
condensates \cite{LMP}.

Our aim is to obtain an effective Lagrangian for a dynamical composite field
$\widehat{S}\left( x\right) $ through radiation corrections to the Dirac
Lagrangian with an external field $S\left( x\right) $ . A composite field $
\widehat{S}\left( x\right) $ is defined by vanishing of its renormalisation
constant $z_S\left( \frac \Lambda \mu \right) \rightarrow 0$ when $\mu $
reaches the compositeness scale $\Lambda $ \cite{Hayashi}. We consider composites in
the approximation of fermion loops resulting from integration over fermions
in the path integrals. If $S\left( x\right) $ were elementary, its
renormalisation constant would be $z_S=1+\Delta ,$ where $\Delta $ contains
logarithms.For a composite $\widehat{S}\left( x\right) $ suitable
normalisation leads to $z_S=\Delta $ .

The Lagrangian for fermions interacting with scalar background $S$ and
vector fields $V_\nu =\left( g_V/2i\right) V_\nu ^k\tau _k,A_\nu =\left(
g_A/2i\right) A_\nu ^k\tau _k$ and corresponding path integral are given by

$$L=\overline{\psi }\left( \not D+S\right) \psi $$

$${\bf Z}\left[ S\right] =\int {\it D}\overline{\psi }{\it D}\psi \exp
\{i\int dxL\}\equiv Z_\psi \left[ D+S\right] . \eqno(10)
$$

We consider only fermion loops as a main reason for existence of a dynamical
scalar $\widehat{S}\left( x\right) $ and in the fermion path integral

integrate out fermions between the gauge invariant compositeness scale $
\Lambda $ and a running scale $\mu =\Lambda \exp \left( -\sigma \right) $ ,
$\sigma \geq 0$. An effective action $W_{eff}\left( \widehat{S}\ \right) $
for the composite field $\widehat{S}$ is then

$$
\exp \left( iW_{eff}\left( \widehat{S}\right) \right) ={\bf Z}_\psi \left[
\not D+S;\Lambda \right] /{\bf Z}_\psi \left[ \not D+S;\mu \right]
\eqno(11)
$$

where ${\bf Z}\left[ ...:\Lambda \right] $ means that the path integral is
extended up to the scale $\Lambda $ . Self-consistency of such definition
requires that kinetic energy of $\widehat{S}$ in the compositenes region
should be positive and disappear at the compositeness scale, while the
potential part of $W_{eff}$ should give positive mass.

We consider first the case of one quark generation $u,d$ with common mass $m$
when dynamics of composite pseudoscalar and vector fields is well described
by the Chiral theory and the relevant energy scales -the low energy QCD
region $R_{QCD}$ - are defined by the quark and gluon condensates \cite{Shifman}. This
region can be described also by the quark spectral asymmetry and invariant
cutoff \cite{LMP} . The chiral and scale anomalies dominate strong interaction
physics in $R_{QCD}$ \cite{PhL}. We assume that $R_{QCD}$ is also the
compositeness region $R_S$ for a scalar field $S\left( x\right) $ .

The quark path integral in the Euclidean space is represented in the form

$$
\ln {\bf Z}_\psi \left[ \hat {\not D}\right] =Tr\ln ( \frac{\not
D}{\Lambda })\cdot \Theta \left( \Lambda ^2-\left( \not D-M\right) ^2\right)
\eqno(12)
$$

with parameters $\Lambda $ and $M$ defined in terms of the quark$C_q$ and
gluon $C_g$ condensates and $\Lambda $ identified with the compositeness
scale $\Lambda _S$ ; here $\kappa $ is the normalization scale.

Let us for a moment omit vector field $V_\mu $ from the Dirac operator $\not
D$ .Integrating out fermions between $\Lambda $ and $\mu =\Lambda \exp
\left( -\sigma \right) ,\sigma \geq 0,$ we get the following effective
lagrangian for $\widehat{S}\left( x\right) $ in the Minkowski space
$$
L_{eff}\left( S\right) =\frac{N_c}{8\pi ^2}\sigma tr_f\left( D_\mu \widehat{S
}\right) ^2+3\sigma F_\pi ^2{}tr_f\widehat{S}^2\xi _2-\frac{N_c}{8\pi ^2}
\sigma tr_f\widehat{S}^4
$$

$$-3\sigma C_q\xi _3tr_f\widehat{S}-\sigma \frac{N_c}{2\pi ^2}M\xi _1tr_f
\widehat{S}^3
\eqno(13)   $$

where $D_\mu =\partial _\mu +\left[ V_\mu ,\circ \right] $ is a covariant
derivative, $\xi _n$ describe running of masses and scalar condensate

$$
\xi _n=\frac {1-e^{-n\sigma }}{n\sigma }
\eqno(14)   $$

$F_\pi $ is the pion decay constant, $N_c$ is number of colors, $\Lambda $
and $M$ are expressed in terms of the condensates

$$
C_q=-\frac{N_c}{2\pi ^2}\left( \Lambda ^2M-\frac{M^3}3\right) ;C_g=\frac{3N_c
}{2\pi ^2}\left( 6\Lambda ^2M^2-\Lambda ^4-M^4\right)
\eqno(15)   $$
The gluon condensate $C_g$ does not appear explicitly in $L_{eff}\left(
\widehat{S}\right) $ ; tr$_f$ refers to flavor matrices. Kinetic energy in $
L_{eff}$ is positive for $\sigma \succ 0$ and disappears at the
compositeness scale $\Lambda $ . Main terms of induced potential $\widehat{U}
\left( \widehat{S}\right) $ are presented in the first line; they include $
\widehat{S}^2$ and $\widehat{S}^4$ terms. The term with $\widehat{S}^2$
enters $L_{eff}$ with a wrong sign , so that $\widehat{U}\left( \widehat{S}
\right) $ may have a minimum.

We see that integrating over quarks with the background field $S$ in the
quark lagrangian give rise to an effective lagrangian $L_{eff}\left(
\widehat{S}\right) $ for a dynamical field $\widehat{S}$ with coefficients
defined by quarks.

\vskip 6pt
{\bf Nonlinear representation of composite field}
\vskip 3pt

We consider nonlinear representation of composite field $\widehat{S}\left(
x\right) $ and denote it in this case by $\Phi =\Phi _0\exp \phi ,$ where $
\Phi _0$ is a large field and $\phi =\phi _0+\phi _k\tau _k$ are small
fluctuations, $\tau _k$ are Pauli matrices for isospin.

To study characteristic features of the model we consider first the case
when the spectral asymmetry $M=0$ and ,consequently, the quark condensate is
absent, $C_q=0.$ The effective lagrangian $L^0=L_{eff}\left( M=0\right) $
has the standard form

$$
L^0\left( \Phi \right) =Z_\Phi \frac 12\left( \partial _\nu \phi _a\right)
^2+\mu ^2tr_f\Phi ^2-\frac \lambda 2tr_f\Phi ^4
\eqno(16)   $$

with $a=0,k,$ renormalization constant $Z_\Phi =N_c\sigma \Phi _0^2/2\pi ^2$
, and potential parameters

$$
\mu ^2=3F_\pi ^2\sigma \xi _2,\lambda =\frac{N_c\sigma }{4\pi ^2}
\eqno(17)   $$

The minimum $\Phi _0$ of the potential $U\left( \Phi \right) $ and masses of
fields $\phi _a$ are given by relations

$$
\Phi _0^2=\frac{\mu ^2}\lambda ,m_\phi ^2=4\Phi _0^2
\eqno(18)   $$
which held independently of $\sigma $ . Dynamical quark mass is $m_\psi
=\Phi _0$ , or $m_\phi =2m_\psi $ . This relation is characteristic for many
models \cite{Nambu}. Masses $m_\phi ^2$ run with $\sigma $ due to factor $\xi _2\left(
\sigma \right) $ and decrease for increasing positive $\sigma $ .

Let us now restore $M$ . The effective lagrangian $L^M$ gets two additional
terms

$$
L^M=L^0-3C_q\sigma \xi _3tr\Phi -\frac{N_cM}{2\pi ^2}\sigma \xi _1tr\Phi ^3
\eqno(19)   $$

and $\Phi _0$ is to be found from the equation

$$
\Phi _0^3+\frac{3N_cM}{4\pi ^2\lambda }\sigma \xi _1\Phi _0^2-\frac{\mu ^2}
\lambda \Phi _0+\frac{3C_q}{2\lambda }\sigma \xi _3=0
\eqno(20)   $$
At the compositeness scale $\sigma =0$ the solution is simple

$$
\left( \Phi _0+M\right) ^2=3\Lambda ^2
\eqno(21)   $$

and in this case $m_\phi ^2=4\left( \Phi _0+M\right) ^2$ while for the quark
dynamical mass we have the same relation $m_\psi =\Phi _0$ as for $M=0$ .
Additional contribution to $m_\phi $ arises from condensates.

In the case $\sigma \not =0$ the mass of scalars $\phi _a$ as a function of $
\sigma $ is defined in terms of $\Phi _0\left( \sigma \right) $ and $\xi $

$$
m_\phi ^2=6\left[ \Phi _0^2+2M\Phi _0\xi _1+\left( M^2-\Lambda ^2\right) \xi
_2\right]
\eqno(22)   $$
$m_\phi $ decreases with increasing positive $\sigma $ from its maximal
value $m_\phi =2\left( \Phi _0+M\right) $ at $\sigma =0$ .

Interaction of scalars $\phi _a$ with pseudoscalar mesons has the same
structure as in the dilaton model for scalars \cite{PhL, VYuN+} but with different
coefficient: $M\xi _1\left( \frac{3\sigma }{2\pi ^2}\right) ^{1/2}$ instead
of $F_\pi .$ It reduces width $\Gamma _\phi $ compared with the dilaton
model.

Let us reintroduce vector fields $V_\nu $ and $A_\nu $ and study their
interaction with scalar condensate $\Phi _0$.The field $V_\nu $ appears only
in kinetic part of $L_{eff}\left( \Phi \right) $ , and $\Phi _0$ does not
break symmetry. Thus, the only impact of integration over quarks on $V_\nu $
is the contribution $\Delta _V=\sigma N_c/12\pi ^2$ to coupling $1/g^2$ in
the original lagrangian $L_V=\left( \frac 1{2g^2}\right) trV_{\mu \nu }^2$
for elementary field $V_\nu $ .

The axial vector $A_\nu $ enters $L_{eff}\left( \Phi ,A_\nu \right) $
quadratically

$$
L_{eff}\left( \Phi ,A_\nu \right) =\frac 23\mu ^2trA_\nu ^2-\frac \lambda
2tr\left\{ A_\nu ,\Phi \right\} ^2-\frac{N_cM}{\pi ^2}\sigma \xi _1tr\left(
A_\nu ^2\Phi \right)
\eqno(23)   $$

It follows that field $A_\nu $ will get contribution $\delta m_A^2=\Delta
_Am_\phi ^2$ to 'bare' mass $m_{0A}^2$ , if $A_\nu $ is elementary, giving
the total mass $m_A^2$

$$
m_A^2=m_{0A}^2+\Delta _Am_\phi ^2
\eqno(24)   $$

where $\Delta _A=\Delta _V=\lambda /3$ is one-loop contribution to the
renormalization constant $Z_A=1+\Delta _A$ , or to $1/g_A^2$ in initial
lagrangian $L_A=\left( 1/2g_A^2\right) tr\langle \left( D_\mu A_\nu -D_\nu
A_\mu \right) ^2+\left[ A_\mu ,A_\nu \right] ^2\rangle $ .The mass $\delta
m_A^2$ induced by scalar condensate depends also on the quark and gluon
condensates through the asymmetry parameter $M.$ , and dissapears at the
compositeness scale $\sigma =0$ . If $A_\nu $ is a composite field with the
same compositeness scale $\Lambda _A=\Lambda _\phi $ , then it acquires the
mass $m_A^2=m_\phi ^2$ .

Let us review results of the model. We have found relations of the scalar
condensate to the quark and gluon condensates and thereby can explain why
mass of $q\overline{q}$ scalar mesons may be considerably higher than two
quark masses $2m_q$ . Certainly, a contribution from gluons is present here,
though composition of scalar meson cannot be identified as $qg\overline{q}$
or $qgg\overline{q}$ etc. If to use this language, on should
rather speak about mixture of combinations $q\left\langle \overline{q}
q\right\rangle _0\overline{q}$ and $q\left\langle G_{\mu \nu
}^2\right\rangle _0\overline{q}$ containing the quark and gluon condensates.

We have found a contribution $\delta m_A^2$ from scalar condensate to the
mass of axial vector meson in the model (a), when internal symmetry is not
broken, so that, for vector mesons, analogous contribution does not exist, $
\delta m_V^2=0$. Thus, both results $m_{q\overline{q}}-2m_q$ and $\delta
m_A^2$ have the same origin in interplay of the quark and gluon
condensates.Another result is that nonleptonic decay widths of scalar mesons
will be lower in this approach than in the case of dilaton-quarkonia \cite{VYuN+}.

\vskip 6pt
{\bf  Conclusions. Relation to other models}
\vskip 4pt

We have demonstrated two ways of choosing collective variables in QFT in order
to find effective actions for composite fields without knowledge of
inessential variables. Both approaches are quite universal if a complex
quantum system is described by functional integral. The group-theoretical
approach leads to an action for composite quantum field, while the Induced
action approach gives us a classical action for composite field. In the
first case collective variables (parameters of group transformation) are not
present in the initial lagrangian. In the second case external fields in the
initial lagrangian have all quantum numbers of required composite fields,
and functional integration provides them with kinetic term and interactions.

Induced action formalism is closely related to that  for a quantum scalar
field, if we assume that the scalar background field in the quark
Lagrangian arises due a mechanism which produces this quantum field with a
scale independent potential U(S) (weight  function) of dim4. If such a
potential is absent, the induced
action for classical scalar field corresponds to the quantum case when the
quark scalar current has zero expectation value in vacuum.
The Nambu-Jona-Lasinio model  \cite{Nambu} contains a dim2 potential U(S). Finite
compositeness scale approach to these models is considered separately  \cite{Nono}.


\begin{thebibliography}{99}
\bibitem{Bethe-Salpeter} H.A.Bethe and E.E.Salpeter, Phys.Rev.82 (1951) 309; ibid 84 (1951) 1232.
\bibitem{Dyson} F.J. Dyson, Phys.Rev. 75 (1949) 1736; J.Schwinger, Proc.Nat.Acad.Sci. 37 (1951) 452.
\bibitem{NN} N.N.Bogolyubov. JETP, 34 (1958), 58,73; N.N. Bogolubov, V.V.Tolmachev, D.V.Shirkov, New
method in the theory of superconductivity,1958,Moscow,
\bibitem{diquark}Yu.V.Novozhilov, A.G.Pronko, D.,V.Vassilevich, Phys.Lett. B343 (1995) 358; B351 (1995) 601,
\bibitem{AAA+YuVN} A.Andrianov and Yu.V.Novozhilov; Phys.Lett. 153B (1985) 422,
A.Andrianov; Phys.Lett. 157B (1985) 425,
\bibitem{PhL} A.A.Andrianov, V.A.Andrianov, V.Yu.Novozhilov, Yu.V.Novozhilov; Phys.Lett.
186B (1987) 401,
\bibitem{Zeld+Sakharov} Ya.B.Zeldovich, Pis'ma Zh.Eksp.Teor.Fiz. 6 (1967) 883,
A.D.Sakharov, Dokl.Acad.Sci.USSR 177 (1967) 70; Teor.Mat.Fiz. 23 (1975) 178,
\bibitem{Adler} S.L.Adler, Rev.Mod.Phys. 54 (1982) 729,
\bibitem{David} F.David, Phys.Lett. B138 (1984) 383,
\bibitem{Grav} Yu.V.Novozhilov, D.V.Vassilevich, Lett.Math.Phys., 21 (1991) 253,
\bibitem{Einst} D.V.Vassilevich, Yu.V.Novozhilov, Vestnik LGU, Physics, 2 (1991) 76,
\bibitem{Fock} V.Yu.Novozhilov and Yu.V.Novozhilov,in Proc. VIII UNESCO Int.School of Physics
 "Quantum Theory in honour of V.Fock", 1998, 220, ed.Yu.Novozhilov, St.Petersburg, 1998.
\bibitem{VYuN} V.Novozhilov; Phys.Lett. 228B (1989) 240,
\bibitem{YuVN} Yu.V.Novozhilov; Phys.Lett. B225 (1989) 165.
\bibitem{Hayashi} K.Hayashi et al; Fortsch.d.Phys. 15 (1967) 625.
\bibitem{LMP} A.A.Andrianov, V.A.Andrianov, V.Yu.Novozhilov, Yu.V.Novozhilov,
Lett.Math.Phys. 11 (1986) 217,
\bibitem{VYuN+}  A.Boduylkov and V.Yu.Novozhilov; Nuovo Cim. 103A (1990) 1381;
    V.Novozhilov, V.Soloviev; Vestnik Sankt-Petersburg University 18 (1993) 26.
\bibitem{Shifman} M.A.Shifman, A.I.Vainstein, V.I.Zakharov; Nucl.Phys. B147 (1979) 385,
\bibitem{Nambu} Y.Nambu, G.Jona-Lasinio; Phys.Rev.122 (1961) 345;
V.Elias,  M.D.Scadron; Phys.Rev.Lett.53 (1984) 1129;   M.K.Volkov; Ann.Phys.(N.Y.) 157 (1984) 282;
 A.Dhar, R.Shankar, S.Wadia; Phys.Rev. D3 (1985) 3256.
\bibitem{Nono} V.Yu.Novozhilov and Yu.V.Novozhilov, "Model with a finite
      compositeness scale for composite scalar  field", Rep. StPbU-104/99.

\end{thebibliography}
\end{document}